\documentclass[12pt,aps,nofootinbib,preprintnumbers]{revtex4}
\usepackage{amsmath,amssymb,graphicx}
\newcommand{\OPhi}{\widehat{\mathcal{O}_{\Phi}}}

\begin{document}
\title{The high momentum behavior of a quark wake}
\author{Amos Yarom}
\affiliation{Ludwig-Maximilians-Universit\"at \\ Department f\"ur Physik, \\ Theresienstrasse 37, 80333 M\"unchen, Germany}
\email{yarom@theorie.physik.uni-muenchen.de}

\preprint{LMU-ASC 09/07}

\begin{abstract}
Using the AdS/CFT duality, we analytically evaluate the high momentum (or short distance) behavior of the color field strength due to a moving quark in an $\mathcal{N}=4$ SYM plasma. We find a fireball-like behavior in the near quark vicinity as predicted in the literature on general grounds, in the context of heavy ion collisions. Our approach to analytically solving the problem is based on a WKB approximation which may be extended to other setups as well.
\end{abstract}

\maketitle

\section{Introduction, summary and discussion}
Recent results from the Relativistic Heavy Ion Collider (RHIC) at the Brookhaven National Laboratory \cite{BRAHMS,PHENIX,PHOBOS,STAR} (see \cite{RHICreview} for a review) have lead to an increasing interest in the details of the strong interactions at temperatures not too far above the deconfinement transition $T_{QCD} \sim 170$~MeV. At these temperatures the gauge coupling $g_{YM}^2$ is, apparently, large. One approach towards understanding this type of strongly interacting phenomenon is through the AdS/CFT duality, where one considers the high temperature phase of large $N$ $SU(N)$ $\mathcal{N}=4$ Super Yang Mills (SYM) theory as a first step towards understanding Quantum Chromodynamics (QCD). While it is not completely clear if, how or when these two theories may be likened \cite{Gubsercomparison}, an ad hoc comparison of some of their features gives remarkably similar results. This includes an analysis of the energy density at strong coupling, the viscosity to entropy density ratio \cite{etatos1,etatos2,etatos3,etatos4}, the diffusion or friction coefficient for quarks and mesons \cite{Washington,Gubserdrag,diff1,drag1,drag2,drag3,drag4,drag5,drag6}, an analysis of meson melting \cite{meson1,meson2}, the jet quenching parameter \cite{jq1,jq2,jq3,jq4,jq5,jq6,jq7,drag7}, the screening length \cite{sl1,sl2} and directional emission \cite{Gubserdilaton,cpotentialdilaton,Gubseremtensor}. See also \cite{other1,other2,other3,other4,other5} for other properties of $\mathcal{N}=4$ SYM which were discussed in the literature in this context.

Being motivated by the relative success of the above analyses, we have considered the directional emission of colored fields due to the motion of an infinitely heavy quark in an $\mathcal{N}=4$ SYM plasma. As the quark moves through the plasma it emits gluons which may be characterized by their directionality and energy. A numerical analysis of such a gluon wake was initiated in \cite{Gubserdilaton}, and continued in \cite{cpotentialdilaton} and \cite{Gubseremtensor} where a comparison to RHIC results was elaborated on. In this note we continue this investigation and, using a WKB\footnote{Wentzel-Kramers-Brillouin-Jeffreys, see for example \cite{WKBbook}.} based approach, obtain an analytic result for the directional emission of colored fields at large momentum scales.

The magnitude of the field strength in response to the motion of the quark, $\text{Tr} F^2$, may be studied by considering the related, Fourier transformed (hatted) quantity $\langle \OPhi(k_-,k_{\bot}) \rangle = -\frac{1}{2 g_{YM}^2} \text{Tr}\left(\widehat{F^2}+ \substack{\text{\tiny{total}} \\ \text{\tiny{derivative}} \\ \text{\tiny{terms}}}\right)$ to be defined in detail in section \ref{S:Calculation}. Here $k_-$ is the momentum component conjugate to $x_{-}\equiv x-vt$, where $x$ is the direction of motion of the quark and $v$ is its velocity, and $k_{\bot}$ is the momentum conjugate to the directions perpendicular to the motion of the quark. We find
\begin{multline}
\label{E:valueofO}
 	\langle \OPhi(k_-,k_{\bot}) \rangle =
 	-\frac{\sqrt{\lambda_{YM}(1-v^2)}}{16}\sqrt{(1-v^2)k_-^2+k_{\bot}^2}
	\\
	+i\frac{\sqrt{\lambda_{YM}(1-v^2)}}{3}\frac{\pi v k_- T^2}{(1-v^2)k_-^2+k_{\bot}^2}
	+\mathcal{O}\left(T^4/((1-v^2)k_-^2+k_{\bot}^2)^2\right).
\end{multline}
In equation (\ref{E:valueofO}) $T$ denotes the temperature of the plasma and $\lambda_{YM}$ the 't Hooft coupling constant of the SYM theory. The first term is identical to the near field contribution of a moving quark in a conformal theory at zero temperature \cite{Danielsson,Gubserdilaton}, while the second term is a purely dissipative effect. We shall concentrate on the latter.

\begin{figure}
\begin{center}
\scalebox{0.8}{\includegraphics{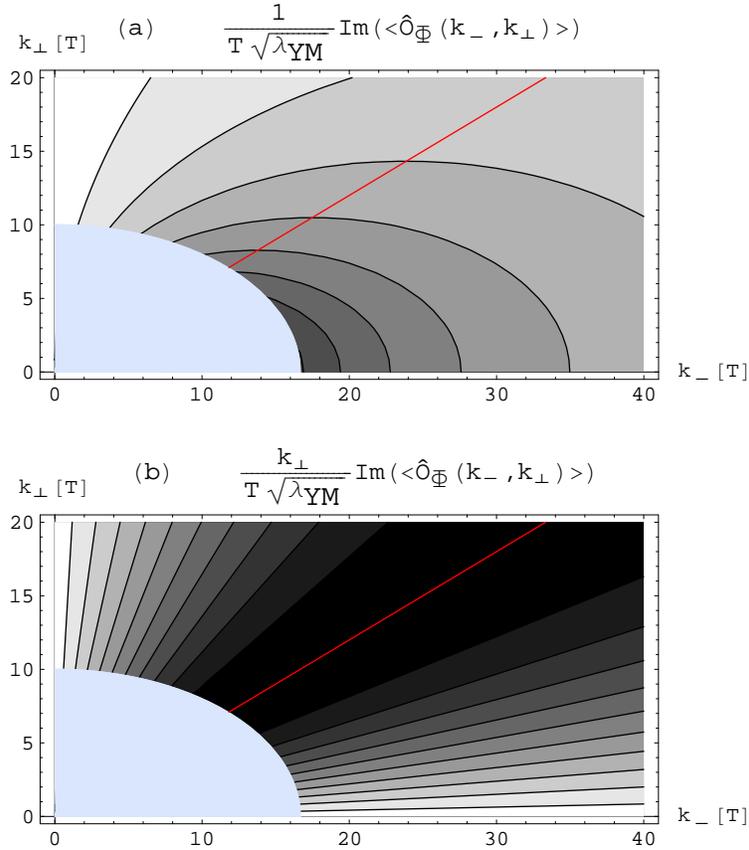}}
\caption{\label{F:momentumspace}Contour plots of the momentum space distribution of the leading dissipative behavior of
the dimensionless quantity $\frac{1}{T\sqrt{\lambda_{YM}}}\langle \OPhi(k_-,k_{\bot}) \rangle = -\frac{1}{T\sqrt{\lambda_{YM}}}\frac{1}{2 g_{YM}^2} \text{Tr}\left(\widehat{F^2}+ \ldots \right)$ in response to a quark moving at a constant velocity of $v=0.8$. The horizontal axis
corresponds to the momentum conjugate to the comoving coordinate of the quark, $k_-$, while the vertical one corresponds to momentum conjugate to the direction transverse to the quark motion, $k_{\bot}$. The momentum is measured in units of temperature (denoted [T] in the plots). Positive imaginary values are shaded dark while imaginary values closer to zero correspond to lighter shading. The origin of momentum space has been colored signifying that this region can not be accessed by our high momentum approximation. Figure (a) depicts the leading dissipative contribution to $\langle \OPhi \rangle$. The red line follows the extremum of the gradient of $\langle \OPhi \rangle$ in the $k_-$ direction. This directional behavior may be graphically enhanced (figure (b)) by multiplying $\langle \OPhi \rangle$ by $k_{\bot}$.}
\end{center}
\end{figure}

A contour plot of the dissipative component of $\langle \OPhi(k_-,k_{\bot}) \rangle$ is given in figure \ref{F:momentumspace}. The directional behavior of this quantity may be analyzed by considering the gradient of $\langle \OPhi \rangle\Big|_{\text{\tiny{d}}}$ (where the `d' implies that we are considering the sub-leading dissipative contribution). In the $k_{\bot}$ direction the gradient of $\langle \OPhi \rangle\Big|_{\text{\tiny{d}}}$ is extremized at $k_{\bot}=0$. In the $k_-$ direction it is extremized at
\begin{equation}
\label{E:ridgeangle}
	\tan\theta \equiv \frac{k_{\bot}}{k_-} = \sqrt{1-v^2}.
\end{equation}
This directional behavior may be graphically enhanced by multiplying the momentum space density $\langle \OPhi \rangle\Big|_{\text{\tiny{d}}}$ by $k_{\bot}$; since $k_{\bot} \langle \OPhi \rangle\Big|_{\text{\tiny{d}}}$ is of order $k^0 \sim 1$ it will not decay at large values of the momenta. The gradient in the $k_-$ direction is unaffected by this amplification, though the gradient in the $k_\bot$ direction is modified reaching a minimum at $k_{\bot}=0$. Topographically, one finds a `ridge' in momentum space at an angle $\theta$ given by equation (\ref{E:ridgeangle}). A comparison of the angle $\theta$ to the numerical value found in \cite{Gubserdilaton} is given in table \ref{T:comparison}.

\begin{table}
\begin{center}
\begin{ruledtabular}
\begin{tabular}{l c c c c}
	velocity					&	0.75	&	0.90	&	0.95	&	0.99 \\
\hline
	Numerical estimation of $\theta$ \cite{Gubserdilaton} ($k_- z_0 = 40$)	&	0.58	&	0.41	&	0.30	&	0.17 \\
	Analytic approximation of $\theta$ (large $k_- z_0$)	&	0.58	&	0.41	&	0.30	&	0.14 \\
\end{tabular}
\end{ruledtabular}
\caption{\label{T:comparison} We compare our results for the angle of the high momentum ridges given in equation (\ref{E:ridgeangle}) to the numerical results presented in \cite{Gubserdilaton}. As the velocity approaches the speed of light, the regime of validity of our approximation shifts towards larger values of the momenta (see section \ref{S:Calculation} for details). The last column shows that for $v=0.99$, the quoted value of $\theta \sim 0.17$ at $\frac{k_-}{\pi T}=40$ \cite{Gubserdilaton} will decrease as $\frac{k_-}{\pi T}$ increases beyond 40, until it will stabilize at an angle of $\theta = 0.14$.}
\end{center}
\end{table}

\begin{figure}
\begin{center}
\includegraphics{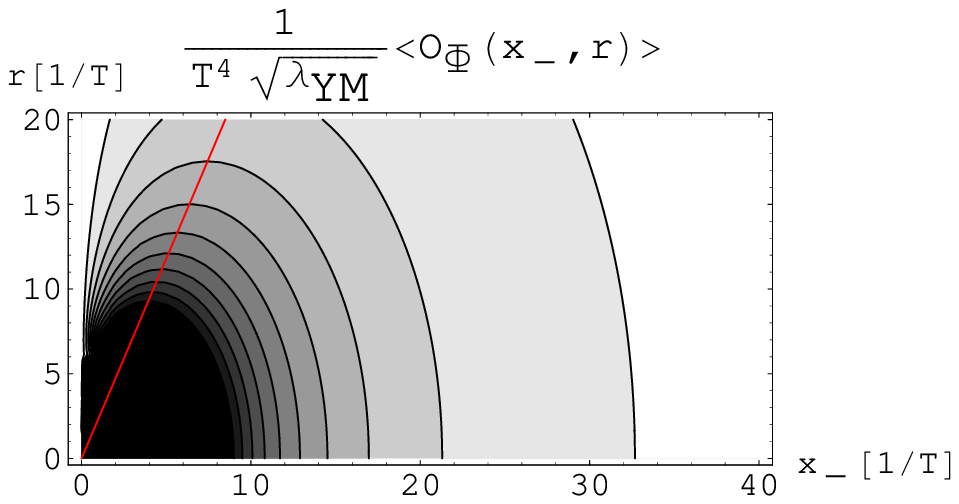}
\caption{\label{F:realspace}Contour plots of the leading dissipative behavior of the dimensionless quantity $\frac{1}{T^4\sqrt{\lambda_{YM}}}\langle {\mathcal{O}}_{\Phi}(x-vt,r) \rangle = -\frac{1}{T^4\sqrt{\lambda_{YM}}}\frac{1}{2 g_{YM}^2} \left(\text{Tr}F^2+ \ldots\right)$ in response to a quark moving at a constant velocity of $v=0.8$. The horizontal axis corresponds to the comoving coordinate of the quark $x-vt$, while the vertical one corresponds to the direction transverse to the quark motion, $r$. Distances are measured in units of inverse temperature (denoted [1/T] in the plot). Large negative values are shaded in dark while values closer to zero correspond to lighter shading. The red line marks the extremum of the gradient in the $x_-$ direction. }
\end{center}
\end{figure}

The momentum space distribution (\ref{E:valueofO}) may be Fourier transformed to real space to obtain the near quark value of the magnitude of the color field strength. One finds that
\begin{equation}
\label{E:valueofOreal}
	\langle\mathcal{O}_{\Phi}\rangle
	=
	\frac{\sqrt{\lambda_{YM}}(1-v^2)^2}{16 \pi^2 \left( r^2(1-v^2)+x_-^2 \right)^2}
    -
    \frac{\sqrt{\lambda_{YM}(1-v^2)}v x_- T^2}{12 \left( r^2(1-v^2)+x_-^2 \right)^{3/2}}
    +\mathcal{O}\left(T^4 (r^2(1-v^2)+x_-^2)^{2} \right).
\end{equation}
Considering again the sub-leading dissipative term, we find that the gradient in the $r$ direction is maximized at $r=0$, while in the $x_-$ direction it is maximized at an angle
\begin{equation}
\label{E:realangle}
	\tan\omega \equiv \frac{r}{x_-} = \frac{\sqrt{2}}{\sqrt{1-v^2}},
\end{equation}
see figure \ref{F:realspace}.

As discussed in \cite{Gubserdilaton,Gubseremtensor}, it is interesting to consider this result in light of the suppression of the back-to-back di-hadron correlations found in the RHIC experiment \cite{btb1,btb2,btb3,btb4}: one expects that jets created as a result of parton collision will be emitted from the quark gluon plasma (QGP) at an angle of $180^\circ$ to each other. Instead, one finds that when considering jets with transverse momenta in the range $2.5 \text{GeV/c} < p_T < 4.0 \text{GeV/c}$, one of the jets is missing or seems to be deflected by an angle of around 1 radian \cite{deflection1,deflection2}. This behavior is interpreted as an indication that the original collision occurred near the surface of the QGP, so that while one of the partons managed to exit the QGP, the other had to traverse it and as a result has been deflected or stopped.

One attempt at explaining this deflection angle is related to the sonic boom created when the quark traverses the QGP at a speed higher than the speed of sound in the plasma \cite{sound1}. In this case the sound wave created far from the moving parton will be directed at a Mach angle relative to the direction of motion of the parton which turns out to be approximately one radian. Indeed, in \cite{Gubseremtensor} it was shown that for a quark moving in an $\mathcal{N}=4$ SYM plasma the low momentum contribution to the energy density has a preferred directional behavior at an angle of $\cos^2\theta \sim \frac{1}{3v^2}$; the Mach angle for a quark in a conformal QGP.

The analysis of the sound wave resulting from the motion of the parton \cite{sound1} takes place in a region far from the jet where perturbations and gradients are small and the hydrodynamic approximation is valid. In contrast, in the near parton region the estimated energy deposition of the parton to the medium is large, probably due to strong dissipative effects. This implies that the jet is surrounded by a ``fireball'' where the hydrodynamic approximation is not valid.

It is interesting to relate this fireball to the behavior of the near field of the parton given in equation (\ref{E:valueofOreal}) and depicted in figure \ref{F:realspace}. The directionality of $\langle \mathcal{O}_{\Phi}\rangle$ (equation (\ref{E:realangle})) indicates that this fireball exhibits Lorentz-like contraction at relativistic velocities.


In order to proceed with an analysis of the short distance behavior of the quark wake, one needs the thermal expectation value of the energy momentum tensor which was obtained numerically in \cite{Gubseremtensor}. The energy momentum tensor will give us a better indication of how energy is distributed by the quark as it travels through the QGP than the field strength. This problem is somewhat more complicated due to the couplings between various components of the (dual) metric tensor's equations of motion. We plan to pursue this in the future.

Finally, we point out that one should be somewhat cautious when relating these high momentum results, which are relevant to $\mathcal{N}=4$ SYM, to real world QCD, especially when considering the high momentum ridges which correspond to a weakly coupled regime of the theory.

\section{\label{S:Calculation}Evaluation of $\langle O_{\Phi} \rangle$.}
In this section we evaluate the thermal expectation value of the operator
$\mathcal{O}_{\Phi} =  -\frac{1}{2 g_{YM}^2} \text{Tr}\left(F^2+ \substack{\text{\tiny{total}} \\ \text{\tiny{derivative}} \\ \text{\tiny{terms}}}\right)$ in the presence of a moving quark in a thermal plasma.

Using the AdS/CFT dictionary \cite{AdSCFT,Witten9802,Gubseretal,MaldacenaWilson}, the motion of a quark moving at constant velocity in the $x$ direction is given by a string in an AdS-Schwarzschild (AdS-SS) geometry whose endpoint is moving with constant velocity $v$ along the asymptotically AdS boundary. The profile of such a string can be found by minimizing the Nambu-Goto action,
\begin{equation}
\label{E:SNG}
	S_{NG}=-\frac{1}{2\pi\alpha^{\prime}} \int e^{\frac{\Phi}{2}}\sqrt{|-h|} d\tau d\sigma.
\end{equation}
In equation (\ref{E:SNG}) we used $h_{\alpha\beta} = G_{\mu\nu}\partial_{\alpha}X^{\mu}\partial_{\beta}X^{\nu}$ with $G$ the AdS-SS metric tensor which may be read off from the line element
\begin{equation}
\label{E:AdSSSline}
	ds^2 = \frac{L^2}{z^2}\left(-g(z) dt^2 + dx^2 + \sum_{i=2}^3 dx_i^2 +\frac{dz^2}{g(z)}\right)
\end{equation}
with $g=1-\left(\frac{z}{z_0}\right)^4$. In this notation, the radial coordinate $z$ lies in the range $0 \leq z \leq z_0$, with $z_0$ the location of the black hole horizon.

Solving for $X^{\mu}$ \cite{Washington,Gubserdrag}, one finds that the string configuration is given by $X = \left(t,\xi\left(z\right)+v t,0,0,z\right)$ with
\begin{equation}
\label{E:defxi}
    \xi(z)=	\frac{v z_0}{4}\left(\ln \frac{1-\frac{z}{z_0}}{1+\frac{z}{z_0}}+2\arctan\frac{z}{z_0}\right).
\end{equation}
(See also \cite{drag1,drag2} for some more elaborate configurations which include the presence of other fields or a chemical potential.)

The AdS/CFT duality then tells us that the thermal expectation value of the CFT operator $\langle \mathcal{O}_{\Phi} \rangle $ is given by
\begin{equation}
\label{E:OandPhi}
	\langle \mathcal{O}_{\Phi} \rangle = -\frac{L^3}{16\pi G_5} \lim_{z \to 0}
		\frac{1}{z^3} \partial_z \Phi,
\end{equation}
where $\Phi$ is the dilaton field in AdS${}_5$ which is sourced by the moving string and $G_5$ is the five dimensional gravitational constant (see equation (\ref{E:Sdilaton}) below). In equation (\ref{E:OandPhi}) one should use
\begin{equation}
\label{E:AdSCFT}
	N^2 = \frac{\pi L^3}{2 G_5},\quad
	\frac{1}{\pi T} = z_0\quad\text{and}\quad
	\sqrt{\lambda_{YM}} = \frac{L^2}{\alpha^{\prime}},
\end{equation}
with $T$ the temperature, $N$ the number of colors and $\lambda_{YM}$ the large $N$ 't~Hooft coupling constant.

The equation of motion for the dilaton ,$\Phi(x-vt,x_2,x_3)$,
may be obtained by minimizing the action
\[
	S=S_{dilaton}+S_{NG}
\]
with
\begin{equation}
\label{E:Sdilaton}
	S_{dilaton}=-\frac{1}{16 \pi G_5} \int \sqrt{|-G|} \frac{1}{2}(\partial \Phi)^2 d^5 x.
\end{equation}
The resulting equation of motion for $\hat{\Phi} = \frac{\alpha^{\prime}L}{4 G_5 \sqrt{1-v^2}}\int \Phi e^{-i k \rho} d^3k$ with $k=(k_-,k_2,k_3)$ and $\rho=(x-vt,x_2,x_3)$ is
\begin{equation}
\label{E:EOMdilaton}
	\left(z^{3}\partial_z \left( z^{-3} g(z) \partial_z\right) -\left(k^2 - \frac{v^2}{g(z)}k_-^2\right)\right)\hat{\Phi}
    	=
    	z e^{-i k_- \xi(z)}.
\end{equation}
In general one finds that when expanding in small $z$, $\hat{\Phi} = M + N z^2 - \frac{1}{3} P z^3 + Q z^4 + \mathcal{O}(z^5)$. The constant $M$ is related by the AdS/CFT duality to a deformation of the Lagrangian (a source term) and in the setup we are considering should be set to zero. This implies also $N=0$. Using (\ref{E:AdSCFT}) and assuming that $P$ depends analytically\footnote{Divergences which are analytic in the momenta are, after a Fourier transform, contact terms and may be ignored (see \cite{Gubseretal,Holren1,Holren2} for general remarks and \cite{ABY1,ABY2} for some recent applications).} on the momenta $k_-,\,k_2$ and $k_3$, equation (\ref{E:OandPhi}) becomes
\begin{equation}
\label{E:OandPhi2}
	\langle \OPhi \rangle
	=
	-\frac{1}{\pi}\sqrt{\lambda_{YM}(1-v^2)} Q.
\end{equation}

The solution to equation (\ref{E:EOMdilaton}) has been evaluated numerically in \cite{Gubserdilaton} and analytically for small momenta in \cite{Gubseremtensor}. In what follows we shall solve it for $k > T$.

Before doing so, we give a physical explanation of our result. This will be followed by a more elaborate proof. We start by noting that as we take the dimensionless quantity $\sqrt{k^2 - v^2 k_-^2} z_0$ to be very large (or equivalently $g\to1$ or $T\to0$), we obtain the equation of motion for a string hanging straight down in empty AdS space. The solution to this equation is uniquely specified by requiring that it be normalizable and source free (i.e., that it vanishes at the asymptotically AdS boundary). One may guess
\begin{equation}
\label{E:dilatonsolutionAdS}
    \hat{\Phi} =
        \frac{\pi z}{2 \tilde{k}^2} \left(\tilde{k} z(I_2(\tilde{k} z)-L_0(\tilde{k} z))+2 L_1(\tilde{k} z)\right),
\end{equation}
where we have introduced $\tilde{k}^2 = k^2-v^2k_-^2$, and $L_i$ is a modified Struve function of order $i$ \cite{AbramowitzStegun}.
From this one can read off $P=1$ and $Q=\frac{1}{16}\pi \tilde{k}$ so that using (\ref{E:OandPhi2}) one has
\[
	\langle \OPhi\rangle = -\sqrt{\lambda_{YM}(1-v^2)} \frac{1}{16} \tilde{k}.
\]

Next, we note that expanding the equations of motion in inverse powers of $\sqrt{k^2 -v^2 k_-^2} z_0 \equiv \tilde{k} z_0$, the leading contributions to the left hand side of (\ref{E:EOMdilaton}) are of order $(\tilde{k} z_0)^{-4}$. Since $z e^{-i k_- \xi(z)} = z + \frac{i v k_-}{3 z_0^2}z^4 + \mathcal{O}(\tilde{k} z_0)^{-4}$, the leading contribution to the right hand side of (\ref{E:EOMdilaton}) is of order $(\tilde{k} z_0)^{-2}$ . At this order the only solution which does not diverge exponentially in the interior and is not sourced is given by the combination
\[
    \hat{\Phi} =
        \frac{\pi z}{2 \tilde{k}^2} \left(\tilde{k} z(I_2(\tilde{k} z)-L_0(\tilde{k} z))+2 L_1(\tilde{k} z)\right)
	 - \frac{i v k_-}{3 \tilde{k}^2 z_0^2}z^4,
\]
which implies that the sub-leading terms in $\langle \OPhi \rangle$ are precisely those given in equation (\ref{E:valueofO}).

Physically we have the following picture: momenta which are large relative to the temperature are associated with a region of spacetime which is close to the asymptotically AdS boundary. Therefore, if one probes large enough momenta, he or she will not notice that the geometry deviates from empty AdS and an analysis of the moving string will be identical to the one in AdS space. Finite temperature corrections are then encoded in the shape of the stringy source near the AdS boundary and not in the spacetime geometry.

In the above discussion there is one issue that we have glanced over and that should be treated with more care. We have expanded the equation of motion in a power series in $(\tilde{k}z_0)^{-1}$ although the normalizability condition on $\hat{\Phi}$ is imposed at the horizon, located at $z=z_0$. This proves to be a non perturbative statement; once we expand the function $\hat{\Phi}(z_0,\tilde{k}; z)$ in a power series in $(\tilde{k}z_0)^{-1}$, we can no longer correctly evaluate $\hat{\Phi}(z_0,\tilde{k}; z_0)$ without resuming the power series. Nonetheless, despite the somewhat hand-waving nature of this argument we shall see in what follows that the final result is correct.

To solve (\ref{E:EOMdilaton}) in the large $\tilde{k} z_0$ limit, we find it useful to switch to the dimensionless variables\footnote{For comparison, the authors of \cite{Gubserdilaton,Gubseremtensor} used $y = \frac{z}{z_0}$, $K_{1} = k_{-} z_0$ and $K_{\bot} = k_{\bot}z_0$, so that $K_1^2(1-v^2)+K_{\bot}^2 = Z_0^2$.
The numerical solution to $B_K$ of \cite{Gubserdilaton} or $Q_A$ of \cite{Gubseremtensor} can be compared to $Q$ through $B_K=Q_A=z_0 Q\Big|_{\substack{z_0 k_- = K_1 \\ z_0 k_{\bot} = K_{\bot}}}$.}
$Z = \tilde{k} z$, $Z_0 = \tilde{k} z_0$, and define $\alpha = v \frac{k_-}{\tilde{k}}$. Redefining $\tilde{k}^2 \hat{\Phi} = \phi$ one obtains the equation of motion
\begin{equation}
\label{E:EOMdilatonmy}
    \left(Z^{3} \partial_Z \left( Z^{-3} g(Z) \partial_Z \right) - 1 + \alpha^2 \left(\frac{1}{g(Z)}-1\right) \right)\phi
    =
    \frac{Z}{\tilde{k}}e^{-i k_- \xi(Z/\tilde{k})}.
\end{equation}

Our approach to solve this equation is to first solve the appropriate homogeneous equation, once with boundary conditions such that the solution has no source, $\phi_-$, and once with boundary conditions such that the solution has no outgoing modes at the horizon\footnote{In the limit $v\to 0$ there are no incoming or outgoing modes at the horizon. In that case one may require that the solution be normalizable \cite{Glueballmass1}. In the discussion following equation (\ref{E:BesseltoHyper}) we show that the $v \to 0$ limit is continuous.}, $\phi_+$. From these solutions, the Greens function, $G(Z,Z')$, for equation (\ref{E:EOMdilatonmy}) or (\ref{E:EOMdilaton}) may be constructed,
\begin{equation}
\label{E:Greens}
    G(Z,Z') = \frac{1}{W(\phi_-,\phi_+)}\begin{cases}
    		\phi_-(Z^{\prime})\phi_+(Z) & Z>Z^{\prime}\\
	       	\phi_+(Z^{\prime})\phi_-(Z) & Z<Z^{\prime},\\
	    \end{cases}
\end{equation}
where $W(\phi_-,\phi_+) = \phi'_+\phi_- - \phi'_-\phi_+$ is the Wronskian of $\phi_+$ and $\phi_-$. The solution to the inhomogeneous equation for $\hat{\Phi}$ is then given by
\[
	\hat{\Phi}(z) = \tilde{k}^{-3}\int_0^{\tilde{k}z_0} G(z \tilde{k},Z') Z^{\prime} e^{-i k_- \xi(Z'/\tilde{k})} dZ^{\prime}.
\]
This method has been used in \cite{Danielsson} to obtain the spacetime form of (\ref{E:dilatonsolutionAdS}), $\langle \mathcal{O}_{\Phi} \rangle = \frac{\sqrt{\lambda_{YM}}}{16 \rho^4}$, for a motionless string (recall that $\rho=(x_-,x_2,x_3)$).

To find the homogeneous solutions to (\ref{E:EOMdilatonmy}) we shall use the WKB approximation. This method has proved to work well when computing the mass spectrum of glueballs (compare \cite{Glueballmass1,Danielsson} to \cite{Glueballmass2}, or for more recent work \cite{Krasnitz} to \cite{BHM1,BHM2}).

In a quantum mechanical context the WKB approximation is valid provided one is considering a high enough momentum or, conversely, a flat enough potential relative to the momentum scale. Since in this work we are interested in the high momentum asymptotics of the solution, we will make sure that the WKB approximation is valid for large $Z_0$. As $Z_0$ grows smaller we will inevitably reach a regime where this approximation will break down.

To apply the WKB approximation to (\ref{E:EOMdilatonmy}) we must first bring it to Schr\"{o}dinger form. This will usually be possible for an asymptotically AdS geometry: consider an equation of the form
\begin{equation}
\label{E:generaldilaton}
	\left(\mathcal{D}_{AdS}+\mathcal{D}^\prime\right) \phi = J,
\end{equation}
where $\mathcal{D}_{AdS} = \partial_Z^2 - 3 Z^{-1} \partial_Z - 1$ such that $\mathcal{D}_{AdS}\phi=0$ is the Fourier transformed equation of motion of a scalar field in $AdS{}_5$ in appropriate coordinates (see the derivation leading to equation (\ref{E:EOMdilatonmy})). The second term on the left hand side of (\ref{E:generaldilaton}) is given by $\mathcal{D}^{\prime} = a_2(Z) \partial_Z^2 + a_1(Z) \partial_Z + a_0(Z)$ and corresponds to a deviation from empty AdS which vanishes at the AdS boundary $Z = 0$. $J$ is a source term. Equation (\ref{E:EOMdilatonmy}) is of this form with
\begin{align}
	a_2(Z) &= g(Z)-1,
	&a_1(Z) &= g'(Z)-\frac{3}{Z}(g(Z)-1),
	\\
	a_0(Z) &= \alpha^2\left(\frac{1}{g(Z)}-1\right),
	&J(Z) &= \frac{Z}{\tilde{k}}e^{-i k_- \xi(Z/\tilde{k})}.
\end{align}

To bring equation (\ref{E:generaldilaton}) to Schr\"{o}dinger form we define $\phi(Z) = \sqrt{\frac{Z^3}{b(Z)}}\psi(Z)$ with $b(Z) = Z^3\exp{\int^Z\frac{-\frac{3}{x}+a_1(x)}{1+a_2(x)}dx}$. For small values of $Z$ the integral evaluates to $-3\ln Z$ which allows us to set $b(0)=1$. Next we divide equation (\ref{E:generaldilaton}) by $\sqrt{\frac{Z^{3}}{b(Z)}}(1+a_2)$. The equation of motion obtained in this way is a Schr\"{o}dinger equation, $\psi''+V \psi = 0$, with a potential
\begin{equation}
\label{E:Schrpot}
    V(Z)=
        \frac{-\frac{15}{4\,Z^2} + (-1+a_0 - \frac{1}{2}a_1^{\prime})}{1 + a_2}
	+
        \frac{9 a_2 + 6 Z(a_1-a_2')+Z^2 a_1 (2 a_2' - a_1)}{4\,Z^2\,{\left( 1 + a_2 \right) }^2}.
\end{equation}
One can check that as long as $a_2'(0)=a_1'(0)=0$, the Schr\"odinger potential approaches $V_{AdS}(Z) = -1 -\frac{15}{4 Z^2}$ in the $Z \to 0$ limit. The corresponding Schr\"{o}dinger equation may be solved explicitly in terms of Bessel functions (see (\ref{E:regionI}) below).

For the case at hand we find
\begin{equation}
\label{E:VPhi}
	V_{\Phi}(Z)=-\frac{15}{4 Z^2}
		-\frac{1 - (1+\alpha^2)\left(\frac{Z}{Z_0}\right)^4}{\left(1-\left(\frac{Z}{Z_0}\right)^4\right)^2}
	       + \frac{4\left(\frac{Z}{Z_0}\right)^6}{Z_0^2\left(1-\left(\frac{Z}{Z_0}\right)^4\right)^2}
\end{equation}
and $b(Z)=1-\left(\frac{Z}{Z_0}\right)^4$. One can see explicitly that as one takes the BH temperature to zero ($Z_0 \to \infty$) or considers small values of $Z$, the potential asymptotes to $V_{AdS}$.

\begin{figure}
\begin{center}
\scalebox{0.7}{\includegraphics{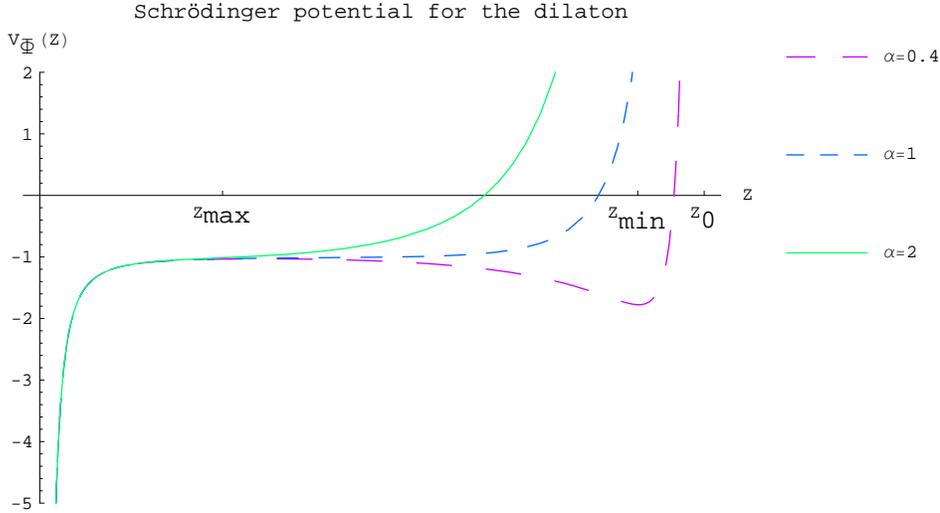}}
\caption{\label{F:PlotVphi} A plot of the Schr\"odinger potential (equation (\ref{E:VPhi})) corresponding to the homogeneous term in the equation of motion for the dilaton (equation (\ref{E:EOMdilatonmy})) for $Z_0 = 40$. As $\alpha$ increases from 0 to 1, the minimum of the potential increases such that both extrema of the potential vanish at $\alpha=1$. For $\alpha>1$, the zero of the potential gets closer to the origin, decreasing the parametric length of the flat region.}
\end{center}
\end{figure}

To get a handle on the functional form of the solutions to the Schr\"odinger equation defined by the potential $V_{\Phi}$ we need to study how $V_{\Phi}$ depends on $\alpha$, see figure \ref{F:PlotVphi}. For $\alpha < 1$ and $Z\in(0,Z_0)$, one finds that it has two extrema which are given by the roots of a polynomial of degree 12. These may be approximated for various values of $\alpha$.

As the potential exits the asymptotically AdS region (valid at small values of $Z$), it reaches a maximum at $Z_{max} = \frac{1}{\sqrt{2}}\left(\frac{15}{1-\alpha^2}\right)^{1/6} Z_0^{2/3}+\mathcal{O}(Z_0^{-2/3})$ where $V_{\Phi}(Z_{max}) = -1 +\mathcal{O}(Z_0^{-4/3})$ after which the potential is approximately constant until it drops sharply to a minimum at $Z_{min} = \left(\frac{1-\alpha^2}{1+\alpha^2}\right)^{1/4}Z_0 + \mathcal{O}(Z_0^{-1})$ where it reaches a value of $V_{\Phi}(Z_{min}) = -\frac{1}{4}\frac{(1+\alpha^2)^2}{\alpha^2}+\mathcal{O}(Z_0^{-2})$. Here we have implicitly assumed that $\alpha \gg Z_0^{-1}$ but $(1-\alpha^2)^{-1/6}$ is small compared to $Z_0^{2/3}$. Other ranges of $\alpha$ may be treated in a similar manner. After passing the minimum, the potential asymptotes to $+\infty$, crossing zero at $\bar{Z} = (1+\alpha^2)^{-1/4}Z_0 +\mathcal{O}(Z_0^{-1})$. With a slight abuse of notation, we shall define $\bar{Z} = (1+\alpha^2)^{-1/4}Z_0$.

For $\alpha \geq 1$ the potential has no extrema. As $\alpha$ increases from 0 to 1 the minimum of the potential, $V_{\Phi}(Z_{min})$, grows closer to $-1$ such that when $\alpha = 1$ the potential is almost flat at intermediate values of $Z$; $V_{\Phi}(Z) = -1 - \frac{15}{4} Z^{-2} + \mathcal{O}\left(\frac{4 Z^6+Z^8}{Z_0^8}\right)$. As one increases $\alpha$ further the value of  $\bar{Z}$ becomes smaller, such that the region where the potential may be approximated as flat becomes shorter.

Since $\alpha$ plays an important role in the shape of the Schr\"{o}dinger potential, we would like to make an aside to discuss the values of $\alpha$ in momentum space. Specifically, we will be studying the value of $\langle \OPhi \rangle$ in the $k_-,k_{\bot}$ plane. Let $\tan\theta = k_{\bot}/k_{-}$. This implies that
\[
	\tan^2{\theta} = v^2-1 + \frac{v^2}{\alpha^2},
\]
so that $0<\alpha^2<\frac{v^2}{1-v^2}$.
Large values of $\alpha$ correspond to small $\theta$ (i.e., $k_{\bot} < k_-$). In order for our large momentum approximation to be valid we will see that $\alpha$ should be restricted by $\alpha < Z_0^{2/3}$. If we wish to probe the small $\theta$ region of momentum space at highly relativistic quark velocities then we will be restricted to large values of $k_-$.

Since the potential $V_{\Phi}$ is technically difficult to deal with, we shall analyze it separately in three different regions, using a different approximation in each one (this was also done in a somewhat different context in \cite{Danielsson}). In the near boundary region we use
\[
	V_1(Z) = -1 - \frac{15}{4Z^2}.
\]
The near horizon region may be approximated by
\[
	V_3(Z) = \frac{\frac{1}{4}Z_0^2 \alpha ^2+1}{4(Z_0-Z)^2} - \frac{Z_0\left(\frac{1}{4}\alpha^2+1\right)+\frac{3}{Z_0}}{4(Z_0-Z)},
\]
and the intermediate region by
\[
	V_2(Z) =  -\frac{1-(1+\alpha^2)\left(\frac{Z}{Z_0}\right)^4}{\left(1-\left(\frac{Z}{Z_0}\right)^4\right)^2}.
\]

One may approximate the regime of validity of these three potentials by ensuring that the correction to $|V_i(Z)-V_{\Phi}(Z)|$ is small. Explicitly, for $V_1$ we require
\begin{equation}
\label{E:defofv1}
	(-1+\alpha^2)\left(\frac{Z}{Z_0}\right)^4 \ll 1,
\end{equation}
$V_2$  is a good approximation in the range
\begin{subequations}
\label{E:defofv2}
\begin{align}
\label{E:defofv2a}
		\sqrt{\frac{15}{4}} & \ll Z, \\
\intertext{and}
        Z_0-Z & \gg \frac{1}{Z_0},
\label{E:defofv2b}
\end{align}
\end{subequations}
(this is a conservative estimate where $\alpha$ is assumed to be small)
and for $V_3$ we find
\begin{equation}
	Z = \mathcal{O}(Z_0).
\end{equation}

Our next step is to solve the appropriate Schr\"{o}dinger equation for the potentials $V_1$, $V_2$ and $V_3$. In each region $i$ we will find two linearly independent solutions which we will denote $\psi^i_a$ and $\psi^i_b$. For region 1 we find
\begin{equation}
\label{E:regionI}
	\psi^{1}_{a} = Z^{1/2} I_2(Z)\quad\text{and}\quad
	\psi^{1}_{b} = Z^{1/2} K_2(Z),
\end{equation}
where here and in what follows we define $I_{\nu}$ and $K_{\nu}$ to be modified Bessel functions of index $\nu$ of the first and second kind.

At small $Z$ we have $\phi \sim Z^{\frac{3}{2}} \psi^1$. A series expansion for small arguments of the modified Bessel functions in equation (\ref{E:regionI}), $I_2(x) \sim x^2$ and $K_2(x) \sim x^{-2}$, then tells us that the solution which does not correspond to a deformation of the Lagrangian (in this case, the solution which vanishes at the AdS boundary) is given by $\psi^{1}_{a}$.

For $V_3$ we have the solutions
\begin{subequations}
\label{E:regionIII}
\begin{align}
\label{E:regionIIIa}
	\psi^{3}_{a} & = -\sqrt{2} \sqrt{Z_0-Z} I_{\frac{i Z_0 \alpha }{2}}\left(\frac{1}{2} \sqrt{Z_0-Z} \sqrt{\frac{\left(\alpha ^2+4\right) Z_0^2+12}{Z_0}}\right),\\
\label{E:regionIIIb}
	\psi^{3}_{b} & = -\sqrt{2} \sqrt{Z_0-Z} K_{\frac{i Z_0 \alpha }{2}}\left(\frac{1}{2} \sqrt{Z_0-Z} \sqrt{\frac{\left(\alpha ^2+4\right) Z_0^2+12}{Z_0}}\right).
\end{align}
\end{subequations}
To understand the near horizon behavior of $\psi^3$ given in equations (\ref{E:regionIII}), we rewrite the modified Bessel functions in terms of Hypergeometric functions
\begin{subequations}
\label{E:BesseltoHyper}
\begin{align}
\label{E:BesseltoHypera}
	I_{\nu}(x) &= \frac{1}{\Gamma(\nu+1)}\left(\frac{x}{2}\right)^{\nu}{}_0F_1\left(\nu+1,\frac{x^2}{4}\right), \\
\label{E:BesseltoHyperb}	
	K_{\nu}(x) & = \frac{1}{2}\Gamma(\nu)\left(\frac{x}{2}\right)^{-\nu}{}_0F_1\left(-\nu+1,\frac{x^2}{4}\right)+
		\frac{1}{2}\Gamma(-\nu)\left(\frac{x}{2}\right)^{\nu}{}_0F_1\left(\nu+1,\frac{x^2}{4}\right),
\end{align}
\end{subequations}
which is valid provided that $\nu = \frac{i}{2} Z_0 \alpha \neq 0$. Defining the Kruskal-like coordinates $Y=\ln(Z_0-Z)$, we find that $(Z_0-Z)^{\pm\frac{1}{2}\nu} = e^{\pm\frac{i}{4} z_0 v k_- Y}$. It is now clear that $I_{\nu}$ corresponds to a solution with modes which are ingoing at the horizon and $K_{\nu}$ to a linear combination of ingoing and outgoing modes. Therefore $\psi_a^3$ satisfies the near horizon boundary conditions we are looking for. For a motionless quark ($\alpha \to 0$), there are no ingoing or outgoing modes at the horizon. Instead one imposes that the solution is normalizable. This implies that we should use the $I_0$ solution, which gives us a nice $v\to0$ limit.

In region 2 we need to use the WKB approximation to solve the equation of motion. Ensuring the validity of the WKB approximation, $\left| \frac{V_2'}{V_2} \right| \ll 1$, implies
\begin{align}
\label{E:AdSdivergence}
	Z &\gtrsim 2 \\
\label{E:Horizondivergence}
	Z &\lesssim Z_0 - 2.
\end{align}
We should also take into account the crossing from $V_2<0$ to $V_2>0$ at $Z \sim \bar{Z}$ if this crossing occurs in region 2.

As long as $V(Z)=0$ at $Z>Z_0-2$ (or, in other words, $\alpha\sqrt{Z_0} \lesssim 1$) the potential $V_2$ is strictly negative. This gives us the solutions
\begin{subequations}
\label{E:regionIIalphasmall}
\begin{align}
\label{E:regionIIalphasmalla}
	\psi^{2}_{a} & = -\frac{1}{\sqrt{2}}
		\left(
		\frac{\left(1-\frac{Z^4}{Z_0^4}\right)^2}{1-\frac{Z^4}{\bar{Z}^4}}
		\right)^{1/4}
		e^{-Z F_1},\\
\label{E:regionIIalphasmallb}
	\psi^{2}_{b} & =
		\frac{1}{\sqrt{2}}
		\left(
			\frac{\left(1-\frac{Z^4}{Z_0^4}\right)^2}{1-\frac{Z^4}{\bar{Z}^4}}
		\right)^{1/4}
	e^{Z F_1},
\end{align}
\end{subequations}
where
\begin{equation}
	F_1(Z) = F_1\left(\frac{1}{4};-\frac{1}{2},1;\frac{5}{4};\left(\frac{Z}{\bar{Z}}\right)^4,\left(\frac{Z}{Z_0}\right)^4\right)
\end{equation}
is the Appell hypergeometric function (see for example \cite{Appell}) which may also be written as a linear combination of Elliptic integrals of the first and third kind. As expected, for $0<Z<Z_0$ this function is real as long as its first argument, $Z/\bar{Z}$, is smaller than unity.

On the other hand, if $\alpha\sqrt{Z_0}>1$ we find that the solutions in region 2 are of the form
\begin{subequations}
\label{E:regionIIalphalarge}
\begin{align}
	\psi^2_a & = -\frac{1}{\sqrt{2}} e^{-\bar{Z} \bar{F}_1}
		\begin{cases}
		      \left(\frac{\left(1-\left(\frac{Z}{Z_0}\right)^4\right)^2}{1-\left(\frac{Z}{\bar{Z}}\right)^4}\right)^{1/4}
				e^{-Z F_1(Z) + \bar{Z} \bar{F}_1}	
				&\substack{V_2(Z)<0 \\ \text{and} \\ |V_2^{\prime}|>|V_2|}\\
		      \sqrt{\pi}\left(\frac{Z_0 \alpha^4}{4(1+\alpha^2)^{9/4}}\right)^{1/6}
		      		B_i\left(\left(\frac{4(1+\alpha^2)^{9/4}}{Z_0 \alpha^4}\right)^{1/3}(\bar{Z}-Z)\right)
				&	|V_2^{\prime}|<|V_2|
		      \\
		      \left(-\frac{\left(1-\left(\frac{Z}{Z_0}\right)^4\right)^2}{1-\left(\frac{Z}{\bar{Z}}\right)^4}\right)^{1/4}
		      \cos\left(-i Z F_1(Z) + i \bar{Z} \bar{F}_1 +\frac{\pi}{4}\right)	&	\substack{V_2(Z)>0 \\ \text{and} \\ |V_2^{\prime}|>|V_2|},
		\end{cases}\\
	\psi^2_b & = \frac{1}{\sqrt{2}} e^{\bar{Z} \bar{F}_1}
		\begin{cases}
		      \left(
		      	\frac{\left(1-\left(\frac{Z}{Z_0}\right)^4\right)^2}
				{1-\left(\frac{Z}{\bar{Z}}\right)^4}
		      \right)^{1/4}
		      e^{+Z F_1(Z) - \bar{Z} \bar{F}_1}	
		      &	\substack{V_2(Z)<0 \\ \text{and} \\ |V_2^{\prime}|>|V_2|}
		\\
		2\sqrt{\pi}\left(\frac{Z_0 \alpha^4}{4(1+\alpha^2)^{9/4}}\right)^{1/6}
		A_i\left(
			\left(\frac{4(1+\alpha^2)^{9/4}}{Z_0 \alpha^4}\right)^{1/3}(\bar{Z}-Z)
		\right)
		&	|V_2^{\prime}|<|V_2|
		\\
		2
		\left(-
		      \frac{\left(1-\left(\frac{Z}{Z_0}\right)^4\right)^2}
			{1-\left(\frac{Z}{\bar{Z}}\right)^4}
		      \right)^{1/4}
		\sin\left(- i Z F_1(Z) + i \bar{Z} \bar{F}_1 + \frac{\pi}{4}\right)	
		& \substack{V_2(Z)>0 \\ \text{and} \\ |V_2^{\prime}|>|V_2|},
		\end{cases}
\end{align}
\end{subequations}
where we define the $Z$ independent constant $\bar{F}_1$ through
\[
	\bar{F}_{1} = F_1(\bar{Z})
		     =\frac{\sqrt{\pi } \Gamma \left(\frac{5}{4}\right)}{2 \Gamma \left(\frac{7}{4}\right)}
		      {}_2F_1\left(\frac{1}{4},1;\frac{7}{4};\frac{1}{(1+\alpha^2)}\right),
\]
and $A_i$ and $B_i$ are Airy functions (see for example \cite{WKBbook}).

In all three regions we have chosen our normalization such that the Wronskian of $\psi_a$ and $\psi_b$ is $-1$.

Our next step is to glue the solutions in each region to one another to obtain a continuous and differentiable function, $\psi$. We do this once with boundary conditions which vanish at the AdS boundary, $\psi_-$, and once with boundary conditions which are ingoing at the horizon, $\psi_+$. The gluing should be done at some intermediate points $Z_{12}$ and $Z_{23}$ corresponding to the transition from $V_1$ to $V_2$ and $V_2$ to $V_3$  respectively, such that in each region $i$ the solution is a linear combination of the form $\psi_\pm = A^i_{a\pm}\psi^i_a+A^i_{b\pm}\psi^i_{b}$ (no sum on $i,a,b$).

We should choose $Z_{12}$ and $Z_{23}$ in the region in which there is overlap between the ranges of validity of the potentials $V_1$ and $V_2$, and $V_2$ and $V_3$. Of course our final result will not depend on the exact value of $Z_{12}$ and $Z_{23}$. We start with $\alpha<1$. Considering $V_1$, we require $Z_{12}\gg \sqrt{\frac{15}{4}}$ and $(-1+\alpha^2)^{1/4} Z_{12}\ll Z_0$. The first constraint comes from requiring that there be an overlap with region 2 (see equations (\ref{E:defofv2a}) and (\ref{E:AdSdivergence})). The second inequality arises by requiring that $V_1$ be a good approximation to $V_{\Phi}$ (equation (\ref{E:defofv1})). Since the maximum of $V_{\Phi}$ is at $Z_{max}$ with $V_{\Phi}(Z_{max}) = -1 + \mathcal{O}(Z_0^{-4/3})$ and since $V_1$ asymptotes to $-1$ at large $Z$, we shall define $Z_{12} = Z_0^{2/3}\sim Z_{max}$.  For region 2 we use $Z_{23} = Z_0-2$ due to equation (\ref{E:Horizondivergence}) which is compatible with (\ref{E:defofv2b}) since $Z_0$ is large. As long as $\alpha$ is small we find that region 1 and region 2 have good overlap. To quantify this statement, we observe that $V_1 = -1 -\frac{15}{4 Z^2} $ and for small $Z$, $V_{2} \to -1 + (\alpha^2-1)\left(\frac{Z}{Z_0}\right)^4$. Our approximation will be good as long as the flat intermediate region is captured by these asymptotics. When $\alpha>1$, the point $\widetilde{Z}$ where the potential is most flat is given by the solution to $V''(\widetilde{Z})=0$,  $\widetilde{Z}=Z_{max}+\mathcal{O}\left(Z_0^{-2/3}\right)$. Since $V(\widetilde{Z})\sim V(Z_0^{2/3}) \sim -1+(\alpha^2 - \frac{19}{4})Z_0^{-4/3}$, requiring that $\alpha < Z_0^{2/3}$ will ensure that there will be an overlap between regions 1 and 2 for $\alpha>1$ as well.

We proceed to evaluate the various $A$'s. Consider for example the requirement that $\psi_-$ be continuous at $Z_{i-1\,i}$
\[
	A_{a-}^{i-1}\psi_a^{i-1}(Z_{i-1\,1})+A_{b-}^{i-1}\psi^{i-1}_b(Z_{i-1\,i}) =
	A_{a-}^i \psi^i_a(Z_{i-1\,i})+A_{b-}^i\psi^i_b(Z_{i-1\,1}).
\]
Similar equations hold for the first derivatives of $\psi_-$ and two more such sets for the $\psi_+$ solution. Together with $A_{b+}^3=A_{b-}^1=0$ and $A_{a+}^3=A_{a-}^1=1$ we have eight sets of equations for the eight remaining coefficients.

The actual values of the $A$'s are listed in appendix \ref{A:As}. Some useful properties of these coefficients are\footnote{The actual values of $A_{b-}^2$, $A_{a+}^1$ and $A_{b+}^2$ are not relevant for the calculation we have in mind and it may be that they are exactly zero. Here we take a more conservative approach which allows for an exponentially damped coefficient.}
\begin{subequations}
\label{E:generalAs}
\begin{align}
\label{E:generalAs2}	
	A_{a+}^1 &\sim e^{-2 Z_{12}}A_{b+}^1,\\
\label{E:generalAs3}	
	A_{b+}^2 &\sim e^{-2 Z_{23}}A_{a+}^1,\\
\label{E:generalAs1}
	A_{b-}^2 & \sim e^{-2 Z_{12}} A_{a-}^2,\\
\label{E:generalAs4}
	A_{b-}^3 &\sim e^{Z_{23}+\ldots},\quad A_{a-}^3\sim e^{Z_{23}-\ldots}.
\end{align}
\end{subequations}
The first three expressions, equations (\ref{E:generalAs1}), (\ref{E:generalAs2}) and (\ref{E:generalAs3}), come about as a result of gluing exponentially increasing and exponentially decreasing solutions: an exponentially decreasing solution must be enhanced by an exponential factor in order to be glued to an exponentially increasing one. Regarding $A_{b-}^3$ in equation (\ref{E:generalAs4}), since $\psi^3_b$ is exponentially decreasing near the AdS boundary (see the complex valued Gamma functions in (\ref{E:BesseltoHyperb})) one must boost it when gluing it to an exponential solution in region 2. Similarly, we will find that $A_{a-}^3 \sim e^{Z_{23}-\ldots}$.

Using (\ref{E:Greens}) and going back to our less explicit notation (equation (\ref{E:generaldilaton})), we find that the solution to (\ref{E:generaldilaton}) is given by
\begin{multline}
\label{E:generalpsi}
    \sqrt{\frac{b(Z)}{Z^3}} \phi = \psi_+(Z) \int_0^Z \frac{J(Z')}{W(Z^{\prime})}\sqrt{\frac{Z'^{3}}{b(Z')}}
    	\psi_-(Z^{\prime}) dZ^{\prime}
	\\
	+\psi_-(Z) \int_Z^{Z_0} \frac{J(Z')}{W(Z^{\prime})}\sqrt{\frac{Z^{'3}}{b(Z')}}\psi_+(Z^{\prime}) dZ^{\prime}.
\end{multline}
with
\begin{align}
	W(Z) &= \frac{Z^3}{b(Z)}
			    \begin{cases}
			 	-A^1_{b+}	&	0<Z<Z_{12}\\
				-A_{a+}^2 A_{b-}^2 + A_{a-}^2A_{b+}^2 & Z_{12}<Z<Z_{23}\\
				-A^3_{b-}	&	Z_{23}<Z<Z_0\\
			    \end{cases}.
\end{align}

Neglecting exponentially suppressed terms, one finds after some algebra that for $Z<Z_{12}$, equation (\ref{E:generalpsi}) reduces to
\begin{multline}
\label{E:generalpsi2}
    	\sqrt{\frac{b(Z)}{Z^3}} \phi = -Z^{1/2} K_2(Z) \int_0^Z \frac{J(x)\sqrt{b(x)}}{x} I_2(x) dx^{\prime}
    \\
    	-Z^{1/2}I_2(Z) \int_Z^{Z_{12}} \frac{J(x)\sqrt{b(x)}}{x} K_2(x) dx^{\prime}.
\end{multline}
Expanding the right hand side of equation (\ref{E:generalpsi2}) in a small $Z$ power series, and recalling that we have defined $b(0)=1$, one finds
\begin{multline}
	Z^{1/2}\left(-\frac{1}{8} J(0)
    		-\left(\frac{J(0) b'(0)}{24}+\frac{1}{12} J'(0)\right) Z
    		+J(0)(\ldots) Z^{2}
        \right)
	\\	
	-Z^{1/2} \left( \frac{1}{8} Z^{2} + \frac{1}{96} Z^{4} \right)
		\int_Z^{Z_{12}}
			\frac{J(x) \sqrt{b(x)}}{x} K_2(x)
	        dx
	+\mathcal{O}(Z^{7/2}).
\end{multline}
The last integral may be evaluated by noting that $K_2$ is exponentially suppressed when its argument is large \cite{AbramowitzStegun}. Therefore, as long as $J \sqrt{b}$ is not singular in the region of integration, one may probe the leading behavior of equation (\ref{E:generalpsi2}) by expanding it in a power series and extending the region of integration to infinity. This will give us
\begin{multline}
\label{E:phiexp1}
	\phi
		 = -\frac{1}{4}J(0)Z^2
		   -\left(\frac{1}{24}J(0)b'(0) + \frac{1}{3}J^{\prime}(0)\right)Z^3
		   \\
		   -\left(\frac{1}{8}C \tilde{k}^{-2} -\frac{1}{6}b'(0)J'(0) +J(0)(\ldots)\right)Z^4
		   -\frac{1}{2}J(0)Z^4 \ln\left(\frac{1}{2}Z\right)
		   +\mathcal{O}(Z^5),
\end{multline}
where we have used
\begin{align}
	\lim_{x\to 0} \int_x^{\infty} x^{-1} K_2(x) dx &= \frac{1}{x^2} +
		\frac{1}{4} \left(2 \ln \left(\frac{1}{2}x\right)+2 \gamma -1\right), \\
	\lim_{x\to 0} \int_x^{\infty} x^{0} K_2(x) dx &= \frac{2}{x} - \frac{1}{2}\pi, \\
	\lim_{x\to 0} \int_x^{\infty} x K_2(x) dx &= -2\ln\left(\frac{x}{2}\right)+1-2\gamma,\\
	\int_0^{\infty} x^{n} K_2(x) dx &= C_n, \quad \text{for }n \geq 2,
\end{align}
where the $C_n$'s are real numbers. For example
\begin{equation}
		C_2 = \frac{3}{2}\pi,\quad C_3 = 8.
\label{E:intofxK2}
\end{equation}
The constant $C$ in equation (\ref{E:phiexp1}) is given by the $Z$ independent term in
\begin{equation}
\label{E:Cintegral}
	\tilde{k}^2\int_Z^{Z_{12}} \frac{J(x) \sqrt{b(x)}}{x}  K_2(x) dx
\end{equation}
where corrections of order $e^{-Z_{12}}$ are neglected.

It is interesting to note that, if $J(0) \neq 0$, divergent terms in the $z \to 0$ limit of $z^{-3}\partial_{z}\hat{\Phi} = \tilde{k}^{2} Z^{-3}\partial_{Z} \phi(Z)$ are, due to the logarithm in equation (\ref{E:phiexp1}), non contact terms and a more careful renormalization is required \cite{Holren1,Holren2}. This may have some impact on the mesonic solutions found in \cite{jq7} where the string lies entirely along the boundary.

It should be emphasized that equations (\ref{E:generalpsi2}) and (\ref{E:Cintegral}) have been obtained via the WKB approximation which was shown to be valid as long as $Z_0>>1$. As we discussed earlier, when considering progressively smaller corrections to $\langle \mathcal{O}_{\Phi} \rangle$ the WKB approximation will eventually break down rendering (\ref{E:Cintegral}) invalid.

For the case at hand we have $J(0)=0,\, J'(0) = \tilde{k}^{-1}$ and $b'(0)=0$. Going back to our original variable $z$ and original function $\hat{\Phi}$, we find
\[
	\hat{\Phi}(z)
		 = -\frac{1}{3}z^3
		   -\frac{1}{8}C z^4 + \ldots.
\]
By our assumptions we find that the leading contribution to the integral
(\ref{E:Cintegral}) in the $\tilde{k}z_0 \to \infty$ limit is
\begin{equation}
\label{E:leadingC}
	\tilde{k} \int_{z \tilde{k}}^{\infty} K_2(x) dx = \frac{2}{z} - \frac{1}{2}\pi\tilde{k} .
\end{equation}
As discussed earlier, the $z$ independent term, $-\frac{1}{2}\pi\tilde{k}$, is the contribution of the near field value and is not related to dissipative dynamics (see equation (\ref{E:dilatonsolutionAdS}) and the analysis which follows it). Using the definition of $C$ via (\ref{E:Cintegral}) in equation (\ref{E:phiexp1}) and $b'(0)=0$, allows us to write
\begin{equation}
\label{E:Formulamain}
	\frac{1}{\sqrt{\lambda_{YM}(1-v^2)}}\langle \OPhi \rangle
	=
	-\frac{1}{16}\tilde{k}
	+\frac{\tilde{k}}{8\pi}\int_0^{(z_0 \tilde{k})^{2/3}}K_2(x)\left(\frac{J(x) \tilde{k} \sqrt{b(x)}}{x} - 1 \right) dx
\end{equation}
where it is understood that terms of order $e^{-(z_0\tilde{k})^{2/3}}$ should be consistently neglected.

Inserting $J(x)=\frac{x}{\tilde{k}}e^{-i k_-\xi(x/\tilde{k})}$ and $b=1-\left(\frac{x}{z_0\tilde{k}}\right)^4$, we find
\begin{equation}
\label{E:onemorestep}
	\frac{1}{\sqrt{\lambda_{YM}(1-v^2)}}\langle \OPhi \rangle =
	-\frac{1}{16}\tilde{k}
	+\frac{\tilde{k}}{8\pi}\int_0^{(z_0 \tilde{k})^{2/3}} K_2 \sqrt{1-\left(\frac{x}{z_0\tilde{k}}\right)^4}
		\left( e^{-i k_- \xi(x/\tilde{k})}-1\right) dx.
\end{equation}
The square root in equation (\ref{E:onemorestep}) may be expanded in a Taylor series with an appropriate remainder which will give an indication of the validity of the expansion. The oscillatory term in the integral may also be expanded in a power series as long as its period is large enough, so that the exponential behavior of $K_2$ enforces the product $K_2 e^{-i k_- \xi}$ to be damped. These requirements imply
\begin{equation}
	\frac{\alpha}{3 Z_0^2} 2^3 \lesssim \frac{\pi}{4}
	\quad
	\text{and}
	\quad
	Z_0 \gtrsim 10.
\end{equation}
Since we are working in the $\alpha < Z_0^{2/3}$ and large $Z_0$ regime, this expansion is always valid. Using equation (\ref{E:intofxK2}) one may approximate the integral in equation (\ref{E:onemorestep}) to obtain equation (\ref{E:valueofO}) as advertised.

\section{Acknowledgements}
I would like to thank J. Friess, J. Grosse, S. Gubser, D. Mateos, G. Michalogiorgakis, S. Pufu and U. Wiedemann for useful discussions and correspondence. I would also like to thank J. Erdmenger, D. Krefl and especially M. Haack for initial collaboration on this project. I am supported in part by the German Science Foundation and by a Minerva fellowship.

\begin{appendix}
\section{\label{A:As} Continuity and differentiability of the solutions}
In section \ref{S:Calculation} we have discussed how the gluing of the solutions $\psi^i_{a/b}$, valid in various ranges of the $Z$ coordinate, may be parameterized by the sets $\{ A_- \}$ and $\{ A_+ \}$. In what follows we give explicitly the large $Z_0$ asymptotics of $\{A_{\pm}\}$ when $\alpha<Z_0^{2/3}$. We separate our analysis into two regions. Consider first values of $\alpha$ which are small relative to $\sqrt{Z_0}$,
$\alpha\sqrt{Z_0} = \epsilon \ll 1$, we find
\begin{align}
	A^2_{a-}&=\frac{11}{16 \sqrt{\pi}}Z_0^{-4/3} e^{2 Z_0^{2/3}}+\mathcal{O}(\epsilon^2)\\
	A^2_{b-}&=\frac{1}{\sqrt{\pi}}\\
	A^3_{a-}&=\frac{25}{32 Z_0}e^{Z_0 K(-1)-2 \sqrt{2} \sqrt{Z_0}}+\frac{11 }{16 Z_0^{4/3}}e^{2 Z_0^{2/3}-Z_0 K(-1)}+\mathcal{O}(\epsilon^2)\\
\notag
	A^3_{b-}&=\frac{275 }{512 \pi  Z_0^{7/3}}e^{-Z_0 K(-1) + 2 Z_0^{2/3}+2 \sqrt{2 Z_0}}+\frac{e^{Z_0 K(-1)}}{\pi }\\
	  &\phantom{=}+\left(-\frac{25 i }{64 \sqrt{Z_0}}e^{Z_0 K(-1)-2 \sqrt{2 Z_0}}-\frac{11 i }{32 Z_0^{5/6}}e^{2 Z_0^{2/3}-Z_0 K(-1)}\right) \epsilon +\mathcal{O}(\epsilon^2)
\end{align}
and
\begin{align}
   	A^2_{a+}&=e^{Z_0 K(-1)+ \mathcal{O}(\epsilon^2)} \left(\frac{1}{\sqrt{\pi }}-i\frac{25 \sqrt{\pi }\epsilon }{64 \sqrt{Z_0}}e^{-2 \sqrt{2 Z_0}}  + \mathcal{O}(\epsilon^2)\right)\\
	A^2_{b+}&=-\frac{25 }{32 \sqrt{\pi } Z_0}e^{2 \sqrt{2 Z_0}-Z_0 K(-1)}+\frac{1}{2} i e^{-Z_0 K(-1)} \sqrt{\pi } \sqrt{Z_0} \epsilon +\mathcal{O}(\epsilon^2)\\
\notag
	A^1_{a+}&=-\frac{25 }{32 Z_0}e^{2
        \sqrt{2 Z_0}-Z_0 K(-1)}-\frac{11 }{16 Z_0^{4/3}}e^{Z_0 K(-1)-2 Z_0^{2/3}}\\
	  &\phantom{=}+\left(\frac{1}{2} i  \pi  \sqrt{Z_0}e^{-Z_0 K(-1)}+\frac{275 i  \pi }{1024 Z_0^{11/6}}e^{K(-1) Z_0-2 Z_0^{2/3}-2 \sqrt{2 Z_0}}\right) \epsilon +\mathcal{O}(\epsilon^2)\\
\notag
	A^1_{b+}&=\frac{275 }{512 \pi  Z_0^{7/3}}e^{-K(-1) Z_0+2Z_0^{2/3}+2 \sqrt{2 Z_0}}-\frac{e^{Z_0 K(-1)}}{\pi }\\
	  &\phantom{=}+\left(\frac{25 i }{64 \sqrt{Z_0}}e^{Z_0 K(-1)-2 \sqrt{2 Z_0}}-\frac{11 i }{32 Z_0^{5/6}}e^{2 Z_0^{2/3}-Z_0 K(-1)}\right) \epsilon
   	+\mathcal{O}(\epsilon^2).
\end{align}
Here $K(-1)$ is the complete elliptic integral of the first kind \cite{AbramowitzStegun}, $K(-1) = \sqrt{\pi}\frac{\Gamma(5/4)}{\Gamma(3/4)}\sim 1.3$
One may check that the $\alpha Z_0^{-1/2} = \epsilon \to 0$ limit coincides with a motionless quark as it should (recall that $\alpha \propto v$). In fact, the analysis was somewhat simplified since the large $Z_0$ behavior of the solutions, $\psi^i$, at the gluing points $Z_{12}$ and $Z_{23}$ has been expanded in a power series in $\epsilon$ around the $v=0$, large $Z_0$ behavior.

In contrast, the large $\alpha$ case, $\alpha > Z_0^{-1/2}$, is somewhat more difficult since a large $Z_0$ expansion requires us to know the asymptotic behavior of complex index Bessel functions where both the absolute value of the index and the argument is large. Considering equations (\ref{E:BesseltoHyper}) this reduces to finding the asymptotic expansion of the hypergeometric function ${}_0F_1$ for large values of both its arguments. Finding the asymptotic behavior of hypergeometric functions is a notoriously difficult problem. Luckily, the analysis is simplified in the limits we are considering and the relevant asymptotic expansion can be carried out, see appendix \ref{A:Asymptoticexpansion}. As long as $\alpha \gg Z_0^{-1/3}$ we find
\begin{align}
	A^2_{a-}&=\frac{ 4 \alpha ^2 + 11 }{16 \sqrt{\pi } Z_0^{4/3}}e^{2 Z_0^{2/3}},
	&A^2_{b-}&=\frac{1}{\sqrt{\pi}},\\
	A^3_{a-}&=-\frac{e^{\bar{Z}\bar{F}_1 -\frac{\pi  Z_0 \alpha }{4}}}{\sqrt{\pi }},
	&A^3_{b-}&=Oe^{\bar{Z}\bar{F}_1 +\frac{\pi  Z_0 \alpha }{4}},\\
	A^2_{a+}&=Oe^{\bar{Z} \bar{F}_1 +\frac{\pi  Z_0 \alpha }{4}},
	&A^2_{b+}&=e^{- \bar{Z} \bar{F}_1 + \frac{\pi  Z_0 \alpha }{4}},\\
	A^1_{a+}&=O \frac{ \left(4 \alpha ^2+11\right)}{16 Z_0^{4/3}} e^{\bar{Z} \bar{F}_1 + \frac{\pi  Z_0 \alpha }{4}-2 Z_0^{2/3}},
	&A^1_{b+}&=O e^{\bar{Z} \bar{F}_1 +\frac{\pi  Z_0 \alpha }{4}}.
\end{align}
where $O$ is a complex oscillating term of order one, which differs slightly in each of the expressions.

To use the approximation in appendix \ref{A:Asymptoticexpansion} (equation (\ref{E:approx0f1b})) we must restrict ourselves to $\alpha \gg Z_0^{-1/3}$. Hence, we are missing the coefficients $\{A_{\pm}\}$ for $Z_0^{-\frac{1}{2}} \sim \alpha$. Nonetheless, since the potential is smooth in $\alpha$ and exhibits no distinctive features for this region of parameter space, then from continuity and the general arguments given in section \ref{S:Calculation} we do not expect the $A$'s to behave differently from (\ref{E:generalAs}).

\section{\label{A:Asymptoticexpansion}An approximation for ${}_0 F_1$}
Consider the hypergeometric function $_0F_1(b,x)$ defined by
\[
	_0F_1(b,x) =  \sum_{n=0}^{\infty} \frac{1}{n!}\frac{x^n}{(b)_n}
\]
with $(b)_n = b(b+1)\ldots (b+n-1)$. We wish to find an approximation for $_0F_1(\beta N, x N)$ when $N$ is large. Expanding in a power series and resuming, we find
\begin{align}
\label{E:approx0f1b}
	{}_0F_1(\beta N, x N)
		&=\sum_{n=0}^{\infty} \frac{1}{n!}\frac{x^n N^n}{(\beta N)_n}\\
		&=\sum_{n=0}^{\infty} \frac{1}{n!}\frac{x^n N^n}{\beta^n N^n + \frac{n(n-1)}{2}\beta^{n-1} N^{n-1}} + \mathcal{O}(N^{-2})\\
		&=\sum_{n=0}^{\infty} \frac{1}{n!}\frac{x^n N^n}{\beta^n N^n}\left(1- \frac{1}{N \beta}\frac{n(n-1)}{2}\right) + \mathcal{O}(N^{-2})\\
		&=e^{\frac{x}{\beta}}\left(1-\frac{1}{N\beta}\frac{1}{2}\left(\frac{x}{\beta}\right)^2\right)+\mathcal{O}\left((\beta N)^{-2},\frac{x}{\beta}\right).
\end{align}
Thus
\[
	{}_0 F_1(\beta N, x N) \to e^{\frac{x}{\beta}}
\]
is a good approximation as long as $\left|\frac{x^2}{N \beta^3}\right|\ll 1$. In appendix \ref{A:As} we apply this expansion with parameters
\begin{equation}
	\beta \sim \frac{i \alpha}{2},\quad
	N = Z_0,\quad
	x \sim (Z_0-Z)(\alpha^2+4),
\end{equation}
(see equations (\ref{E:regionIII}) and (\ref{E:BesseltoHyper})). For $Z_0-Z \sim 1$, we find that we must require both $\alpha \gg Z_0^{-1/3}$ and $\alpha \ll Z_0$ in order for the approximation to be valid.
\end{appendix}

\bibliography{JQ}

\end{document}